\documentclass{INTERSPEECH2023}
\usepackage{cite}
\usepackage{amsfonts,amssymb}
\usepackage{amsmath}
\usepackage{bm}
\usepackage{pifont}
\usepackage{multirow}
\newcommand{\cmark}{\ding{51}}%
\newcommand{\xmark}{\ding{55}}%

\interspeechcameraready

\title{Speech Enhancement with Multi-granularity Vector Quantization}
\name{Xiao-Ying Zhao$^1$, Qiu-Shi Zhu$^2$, Jie Zhang$^2$}
\address{
	$^1$University of Science and Technology of China (USTC), Hefei, China\\
	$^2$NERC-SLIP, University of Science and Technology of China (USTC), Hefei, China}
\email{xyzhao1123@mail.ustc.edu.cn, qszhu@mail.ustc.edu.cn, jzhang6@ustc.edu.cn}

\begin{document}

\maketitle
 
\begin{abstract}
With advances in deep learning, neural network based speech enhancement (SE) has developed rapidly in the last decade.
Meanwhile, the self-supervised pre-trained model and vector quantization (VQ) have achieved excellent performance on many speech-related tasks, while they are less explored on SE.
As it was shown in our previous work that utilizing a VQ module to discretize noisy speech representations is beneficial
for speech denoising, in this work we therefore study the impact of using VQ at different layers with different number of codebooks.
Different VQ modules indeed enable to extract multiple-granularity speech features.
Following an attention mechanism, the contextual features extracted by a pre-trained model are fused with the local features extracted by the encoder, such that both global and local information are preserved to reconstruct the enhanced speech.
Experimental results on the Valentini dataset show that the proposed model can improve the SE performance, where the impact of choosing pre-trained models is also revealed.
 
\end{abstract}

\section{Introduction}

Speech enhancement (SE) aims at mitigating background noises and reverberation contained in noisy speech recordings, which is widely configured as a front-end in, e.g., speech recognition~\cite{7403942,zhu2022joint}, assistive hearing~\cite{9664313}, robust speaker recognition~\cite{9064910}.
Traditional statistics-based SE methods usually depend on the stationarity assumption, which, however hardly holds in practice. Recently, the development of deep neural networks (DNNs) facilitates a new direction for performing SE, which even shows a superiority over conventional approaches, particularly in non-stationary scenarios. In general, DNN based SE models can be implemented in a generative~\cite{pascual17_interspeech,michelsanti17_interspeech} or discriminative fashion~\cite{defossez20_interspeech,9746169}. 
The generative models aim to learn the speech distribution, where noisy speech is fed into the model to generate the estimated speech signal subject to a consistency constraint on the distributions of enhanced and clean speech. 
The latter is based on directly minimizing the difference between enhanced and clean signals by training discriminative models, such that the impact of ambient noises on the spectrum can be removed.
The discriminative model can be implemented in the frequency domain~\cite{8659692,hu20g_interspeech} or time domain~\cite{8707065,defossez20_interspeech}, where the frequency- domain SE is achieved by estimating a mask matrix, while the time-domain counterpart is usually implemented following an encoder-decoder fashion.

Discrete representation learning has demonstrated an efficacy in many fields, especially for low-resource tasks,  where vector quantization (VQ) is often used. For example, in~\cite{razavi2019generating,gu2022vector} VQ was used to learn discrete latent representations to generate high quality images. In~\cite{du2022vqtts,liu2022delightfultts}, VQ was used for speech synthesis to generate more natural speech by learning discrete representations. In~\cite{jiang2022cross}, scalable VQ was applied to encode multi-scale features for speech coding.
Considering that learning discrete representations can yield meaningful discrete units and diminish noise components, it was shown in~\cite{zhao2022speech}  that VQ was also applicable to SE by reconstructing speech from discrete noisy representations. However, it was also shown that VQ with a single fixed codebook cannot adequately learn the clustering units at different granularities, which limits its applicability for SE.

In the speech domain, self-supervised pre-training models~\cite{baevski2020wav2vec,zhu2021improved,hsu2021hubert,9747379, baevski2022data2vec,zhu2022robust}, e.g., wav2vec2.0~\cite{baevski2020wav2vec}, HuBERT~\cite{hsu2021hubert}, data2vec~\cite{baevski2022data2vec}, have been rapidly developed by exploiting large amounts of unlabeled data, which were shown to be helpful for various downstream tasks.
The difference therein is that indeed wav2vec2.0, HUBERT and data2vec perform  contrastive learning, masked prediction and regression tasks, respectively.
The SE performance can also be boosted by using self-supervised pre-trained models.
In~\cite{9746303}, thirteen pre-trained models were applied to  SE tasks, which are taken as feature extractors to estimate spectral masks to reconstruct the clean speech waveform. This implies that the high-level features extracted by self-supervised pre-trained models are  applicable to SE, which can also be combined with traditional acoustic features~\cite{hung2022boosting}. 
These pre-trained features are all utilized in the estimation of time-frequency mask, but not validated in the U-Net based SE models.

In this paper, based on~\cite{zhao2022speech}, which uses a single VQ module with a fixed codebook, we therefore explore how VQ in combination with pre-training models can be better leveraged for the SE task.
Considering that different layers of the SE model learn different granularity of information, we first propose a multi-granularity VQ technique to model different levels of information by utilizing different VQ at different layers.
Specifically, different VQ modules have different numbers of codebooks and learnable codevectors.
Compared to~\cite{zhao2022speech}, it is shown that  the performance can be improved due to the use of multi-granularity VQ at the cost of a small increase in the model parameters.
Secondly, to enable the model to better make use of global and local information, we fuse the contextual features extracted by the pre-trained model with the local features extracted by the encoder of the SE model. Both the contextual and local features are input to the U-Net framework.
This differs from the scheme in~\cite{9746303,hung2022boosting}, where only the features extracted by the pre-trained model are incorporated to estimate the time-frequency mask, as the considered features for the proposed method become much richer.
In addition, it is shown that different pre-trained models have a  different  effect on the performance, where the data2vec model performs the best with nearly the same amount of parameters. Experimental results on the Valentini dataset demonstrate that utilizing the multi-granularity VQ in different layers is beneficial for SE. The rest of this paper is organized as follows. Section 2 presents the proposed SE method. Experimental setup and results are shown in Section 3 and Section 4, respectively. Finally, Section 5 concludes this work.

\begin{figure}[!t]
	\centering
	\includegraphics[width=0.48\textwidth]{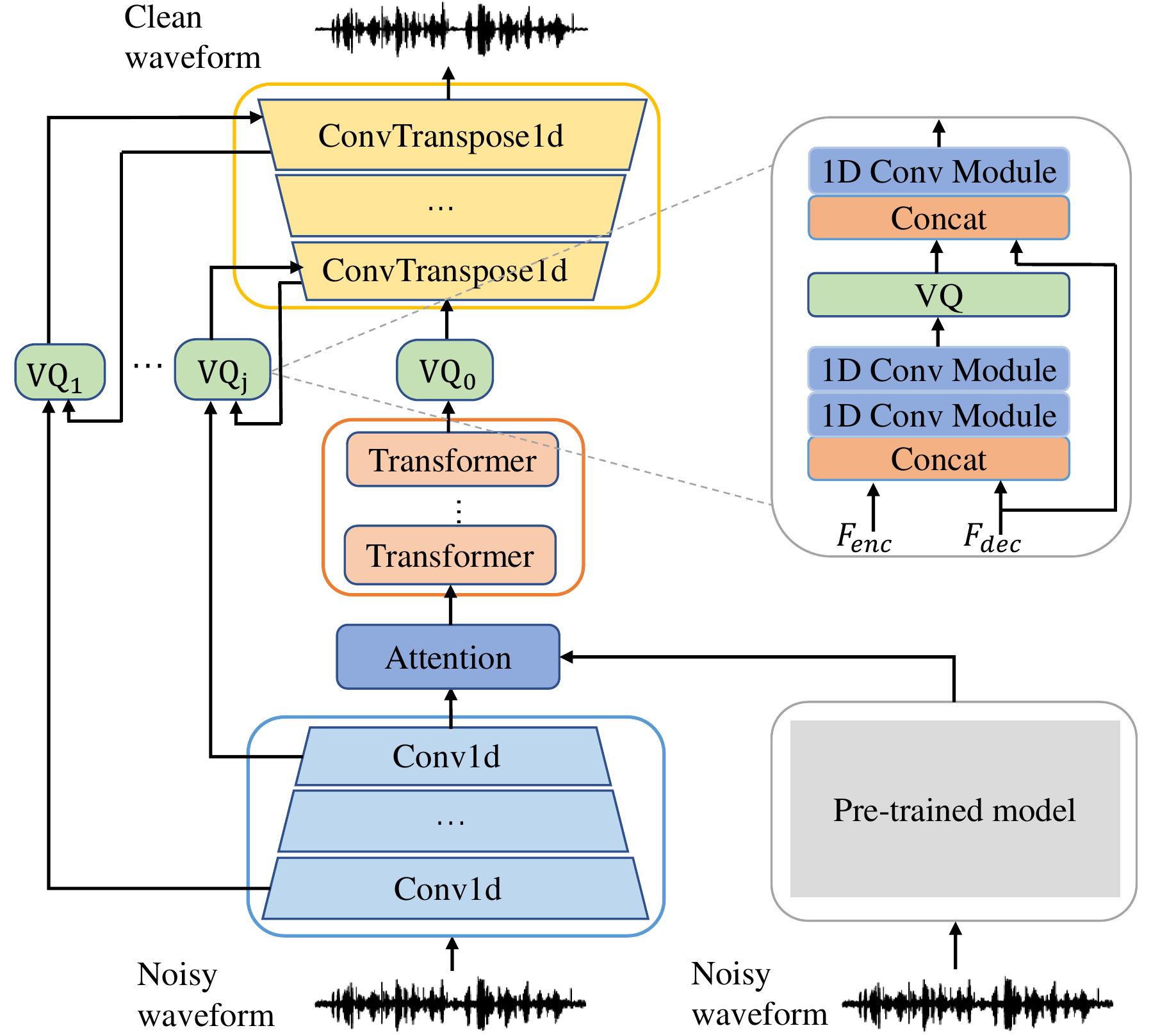}
	\caption{An illustration of the proposed SE model with multi-granularity VQ modules and self-supervised pre-trained model. The feature fusion mechanism of the proposed multi-granularity VQ is shown at the upper right corner.}
	\vspace{-0.2cm}
	\label{fig:fig1}
\end{figure}
\vspace{-0.2cm}
\section{Method}
\label{sec:method}
\vspace{-0.1cm}
To guide the reader, in this section we first briefly introduce the SE model in~\cite{zhao2022speech}, which is the basis of the proposed method.
The SE model follows the U-net structure, which contains convolution-based encoders and decoders and the Transformer encoder based bottleneck layer. This is similar to the DEMUCS network. The VQ module is used to discretize the representations output of the bottleneck layer, which is then fed to the decoder to reconstruct clean speech waveforms, and it is shown in Section 4 that both the VQ module and pre-trained models can be beneficial for the performance. 
However, we  found that a single VQ cannot learn speech representations at different granularities. 
For more details, we refer to~\cite{zhao2022speech}.

\subsection{Model structure}
The structure of our proposed model is shown in Fig.~\ref{fig:fig1}, which mainly contains encoder, bottleneck layer, multi-granularity VQ and decoder.
The encoder and decoder contain $D$-layer one-dimensional (1D) convolution and $D$-layer 1D transposed convolution, respectively, and the bottleneck layer consists of $N$-layer Transformer encoder.
A skip connection is utilized between the $i$-th encoder layer and the $i$-th decoder layer.
The encoder adopts a convolutional module that can efficiently extract local features. To better utilize the pre-trained model to model long-time contextual information, we take the pre-trained model as a feature encoder.
Local features and global features are fed into the attention network for fusion.
The fused features are sent to the bottleneck layer, whose output is fed to the decoder after being discretized by quantizer $VQ_0$.
In order to incorporate discrete units of different granularity, we utilize different vector quantizers.
Specifically, the output of the $i$-th encoder layer and the output of the $i$-th decoder layer are fused into $VQ_i$ to obtain the $i$-th discrete representation, which is then fused with the output of the $i$-th decoder layer. 

\subsection{Multi-granularity VQ}

The proposed multi-granularity VQ mechanism is shown in the upper right corner of  Figure.~\ref{fig:fig1}, which mainly contains the fusion VQ modules.
We first concatenate the output of the $i$-th encoder layer and the output of the $i$-th decoder layer.
The combined features are fed into the two-layer 1D convolution to obtain the fused features,which are input into the VQ module to obtain the discrete features.
Then, the discrete features and the decoder output are concatenated and fed to the 1D convolution module to construct the final output.
The VQ module is built on production quantization, where different number of codebooks and codewords for different layers are considered.
Given $G$ codebooks, each with $V$ learnable $d$-dimensional codewords, we first map the input feature to $\boldsymbol{l} \in \mathbb{R}^{G \times V}$  logits, and then select the discrete vectors by the Gumbel-softmax~\cite{jang2016categorical} operation in a differential way.
The probability of choosing the $v$-th codeword from the $g$-th codebook is given by

\begin{equation}
	\overline{p}_{g,v}=\frac{\exp(\overline{l}_{g,v}+n_{v})/\tau}{\sum_{k=1}^{V}\exp(\overline{l}_{g,k}+n_k)/\tau},
	\label{eq1}
\end{equation}
where $n_v=-\log(-\log(u))$ with $u$  uniformly distributed over [0, 1] and $\tau$ is a non-negative temperature coefficient, 
In the forward stage, the code vector is selected by $argmax$ operation, and in the backward propagation stage the true gradient of the Gumbel-softmax output is utilized.
As it is expected that more codewords can be used for quantization by maximizing the softmax distribution $\bf{\rm l}$, we consider the diversity loss function $L_d$, which is given by
\begin{equation}
	L_{d} = \frac{1}{GV}\sum_{g=1}^{G}\sum_{v=1}^{V}\overline{p}_{g,v}\log \overline{p}_{g,v}.
	\label{eq2}
\end{equation}

\begin{table*}[!t]
	\caption{Performance comparison of objectively evaluation metrics on the Valentini test set.}
	\label{tab:table1}
	\centering
	\begin{tabular}{l|c|ccccc}
		\hline
		\textbf{Model}  & \textbf{Domain}    & \begin{tabular}[c]{@{}c@{}}\textbf{PESQ (WB)}\end{tabular} & \begin{tabular}[c]{@{}c@{}}\textbf{STOI (\%)}\end{tabular} & \begin{tabular}[c]{@{}c@{}}\textbf{pred.} \textbf{CSIG}\end{tabular} & \begin{tabular}[c]{@{}c@{}}\textbf{pred.} \textbf{CBAK}\end{tabular} & \begin{tabular}[c]{@{}c@{}}\textbf{pred.} \textbf{COVL}\end{tabular}  \\ \hline
		Noisy  & -    & 1.97                                                & 92.1                                                & 3.35                                                 & 2.44                                                 & 2.63                                                       \\ \hline
		SEGAN~\cite{Pascual2017}  & waveform     & 2.16                                                & -                                                   & 3.48                                                 & 2.94                                                 & 2.80                                                      \\
		Wave U-Net~\cite{macartney2018improved} & waveform  & 2.40                                                & -                                                   & 3.52                                                 & 3.24                                                 & 2.96                                                      \\
		SEGAN-D~\cite{9201348} & waveform    & 2.39                                                & -                                                   & 3.46                                                 & 3.11                                                 & 3.50                                                      \\
		MMSE-GAN~\cite{8462068}  & time-frequency  & 2.53                                                & 93.0                                                  & 3.80                                                 & 3.12                                                 & 3.14                                                      \\
		MetricGAN~\cite{pmlr-v97-fu19b}  & waveform  & 2.86                                                & -                                                   & 3.99                                                 & 3.18                                                 & 3.42                                                      \\
		DeepMMSE~\cite{9066933}  & waveform  & 2.95                                                & 94.0                                                  & 4.28                                                 & 3.46                                                 & 3.64                                                     \\
		DEMUCS~\cite{defossez20_interspeech}  & waveform    & 3.07                                                & 95.0                                                  & 4.31                                                 & 3.40                                                  & 3.63                                                      \\ 
		CleanUNet~\cite{9746169} & waveform          & 3.09             & 95.8                     & 4.38          & 3.47   & 3.69    \\
		Our previous~\cite{zhao2022speech} & waveform      & 3.13                                                & 96.1                                               & 4.46                                                  & 3.56                                                & 3.82                                                    \\ \hline
		Ours & waveform & 3.18 & 96.7 & 4.53 & 3.64  & 3.85
		\\ \hline
		
	\end{tabular}
\end{table*}

\subsection{The total loss function}
\begin{equation}
	L_{\rm total}=L_{\rm se} + \sum_{i=0}^D\lambda_{i} L_{d_i},
	\label{eq3}
\end{equation}
where $L_{d_i}$ is the codebook diversity loss of the $i$-th VQ and $\lambda_i$ is the weighting parameter, and 
$L_{\rm se}$ is the SE loss function, which is given by~\cite{defossez20_interspeech,zhao2022speech}

\begin{equation}
	L_{\rm se} = \frac{1}{T}  ||\boldsymbol{y} - \boldsymbol{\hat{y}}||_1 + \sum_{i=1}^{M}L_{\rm stft}^{(i)}(\boldsymbol{y}, \boldsymbol{\hat{y}}) ,
	\label{eq4}
\end{equation}
where
\begin{align}
	L_{\rm stft}(\boldsymbol{y},\boldsymbol{\hat{y}}) &= L_{\rm sc}(\boldsymbol{y},\boldsymbol{\hat{y}}) + L_{\rm mag}(\boldsymbol{y},\boldsymbol{\hat{y}}),
	\label{eq5} \\
	L_{\rm sc}(\boldsymbol{y},\boldsymbol{\hat{y}}) &= \frac{|| | {\rm STFT}(\boldsymbol{y})| - | {\rm STFT}(\boldsymbol{\hat{y}})| ||_F}{||  {\rm STFT}(\boldsymbol{y}) ||_F},
	\label{eq6}\\
	L_{\rm mag}(\boldsymbol{y},\boldsymbol{\hat{y}}) &= \frac{1}{T} || \log |{\rm STFT}(\boldsymbol{y})|-\log |{\rm STFT}(\boldsymbol{\hat{y}})|  ||_1,
	\label{eq7}
\end{align}  
where $T$ denotes the speech length, $\boldsymbol{y}$ and $\hat{\boldsymbol{y}}$ are the clean and enhanced speech signals, respectively, $M$ denotes the number of multi-scale STFT loss functions,  $||\cdot||_F$ the Frobenius norm and $||\cdot||_1$ the $L_1$ norm. The multi-resolution loss $L_{\rm stft}^{(i)}$ utilizes the STFT loss with the number of FFT bins ranging from  \{512, 1024, 2048\}, hop size from \{50, 120, 240\} and window length from  \{240, 600, 1200\}, respectively.

\vspace{-0.2cm}
\section{Experimental setup}
\label{sec:experiment}
\subsection{Data description}
To evaluate the effectiveness of the proposed method, we utilize the Valentini~\cite{Valentini2017} dataset, which contains 28.4 hours of clean-noisy speech pairs consisting of 34647 mono audio samples for training and 824 mono audio samples for testing at a sampling rate of 48 kHz. The duration of the audio ranges from 1 second to 15 seconds (the average duration is around 3 seconds). These data are collected from 84 speakers with 10 different types of noises added to the clean speech at 4 signal-to-noise ratios (SNRs) (i.e., 0, 5, 10 and 15 dB) in the training set and 5 types of noise added to the speech with 4 SNRs (2.5, 7.5, 12.5 and 17.5 dB) in the test set, and all speakers are native English. The raw speech waveforms are downsampled to 16kHz and then applied the remix and bandmask techniques for data augmentation as in~\cite{defossez20_interspeech}.

The performance of the proposed method is evaluated using objective metrics, including 1) perceptual evaluation of speech quality (PESQ)~\cite{pesq}, 2) short-time objective intelligibility (STOI)~\cite{5713237}, 3) mean opinion score (MOS) prediction of distortion of speech signal (SIG)~\cite{4389058}, 4) MOS prediction of intrusiveness of background noise (BAK)~\cite{4389058}, 5) MOS prediction of overall quality (OVRL)~\cite{4389058}. 

\subsection{Model configuration}

Our network is built on the basis of the DEMUCS model~\cite{defossez20_interspeech}, which is modified by replacing the LSTM with Transformer and using an attention module as feature fusion module.
Both the encoder and decoder have $D$ = 5 layers and the dimension is 512. We utilize 2 Transformer encoders in the bottleneck layer, where each has a dimension of 512, the feedforward neural network has a dimension of 2048, and the self-attention module has 12 heads.
The 1D convolution in the encoder uses a kernel size of 8 and a stride size of 2.
For the pre-trained models, we use the publicly available wav2vec2.0 base\footnote{https://github.com/facebookresearch/fairseq/tree/main/examples/wav2vec}, HuBERT base\footnote{https://github.com/facebookresearch/fairseq/tree/main/examples/hubert} and data2vec base\footnote{https://github.com/facebookresearch/fairseq/tree/main/examples/data2vec} models in experiments.
All three models are obtained by self-supervised pre-training using the same unlabeled data and have similar parameters.
These model parameters are not used for training, and we only use them to extract features.
For the attention fusion module, we adopt a cross-attention layer with a dimension of 512.
For $VQ_0$ and $VQ_{i}  (i=1, 2, 3, 4, 5)$ modules, we adopt $G$ = 2 codebooks with $V$ = 320 and $G = 1$ codebook with $V$ =  \{320, 640, 960 2560, 5120\} learnable 128-dimensional vectors, respectively.  $\lambda_i$ in (\ref{eq3}) is set to be 0.01 due to the range of the diversity loss, since the amplitude of the codebook diversity loss function $L_d$ is approximately 100 times larger than that of the SE loss function $L_{se}$.
We use the Adam optimizer to train the SE model for 1M iterations, where the batch size is 30 and the maximum learning rate is $2 \times 10^{-4}$. All models are trained on 4 Tesla-V100-32G GPUs.	

\vspace{-0.2cm}
\section{Experimental results}
\label{sec:result}

\textbf{Comparison methods: } We first measure objective evaluation metrics on the noisy test set as a reference. The time-domain comparison methods include SEGAN~\cite{Pascual2017}, SEGAN-D~\cite{9201348} and MetricGAN~\cite{pmlr-v97-fu19b}, which are based on the  generative adversarial network (GAN). The time-frequency domain comparison approaches include MMSE-GAN~\cite{8462068} and DeepMMSE~\cite{9066933}, which estimates the noise spectral density under the minimum mean square error (MMSE) criterion.  Wave U-Net~\cite{macartney2018improved}, DEMUCS~\cite{defossez20_interspeech} and CleanUNet~\cite{9746169} are also compared, which depend on the U-net network in the time domain.

Table~\ref{tab:table1} shows the experimental results of different models on the Valentini dataset.
It can be seen that the proposed method using pre-trained model and multi-granularity VQ can achieve a PESQ of 3.18 and an STOI of 96.7, respectively, which is the best among all comparison methods and is better than our previous method in~\cite{zhao2022speech}, which utilizes a fixed codebook and adopts a pre-trained model to initialize the encoder. This implies that using multi-granularity VQ modules and a pre-trained model as the feature extractor are more effective than the single-fixed counterpart.
In addition, the deployment of multiple codebooks  only results in a small increment of the  parameter amount. (The codebook has codevectors that are learnable, so the number of model parameters added can be calculated as $(320+640+960+2560+5120+320\times2)\times 128/10^6\approx1.3M$.)
In order to more clearly see the impact of each module, we conduct comparative experiments as shown in Table~\ref{tab:table2}.
The base model does not use VQ or pre-trained models, leading to a PESQ of 3.07 and an STOI of 95.2, which are similar to that of CleanUNet, because their model structures are rather similar.
In case the multi-granularity VQ modules are included for the base model as shown in Figure.~\ref{fig:fig1}, the SE performance in all metrics is clearly improved, showing the effectiveness of using multiple codebooks.
By the inclusion of multiple VQs and using any pre-trained model (wav2vec2.0, HuBERT or data2vec) as the feature extractor (i.e., fusing the extracted contextual features with the local features extracted by the SE encoder),  the performance can be further improved to some extent, while  the features extracted from the data2vec model seem the best and the benefits from using wav2vec and HuBERT are almost the same.
\begin{table}[!t]
	\caption{The impact of different modules on the SE performance.}
	\label{tab:table2}
	\centering
	\resizebox{.99\columnwidth}{!}{
	\begin{tabular}{l|lllll}
		\hline
		\textbf{Model}        & \textbf{\begin{tabular}[c]{@{}l@{}}PESQ\\ (WB)\end{tabular}} & \textbf{\begin{tabular}[c]{@{}l@{}}STOI\\ (\%)\end{tabular}} & \textbf{\begin{tabular}[c]{@{}l@{}}pred.\\ CSIG\end{tabular}} & \textbf{\begin{tabular}[c]{@{}l@{}}pred.\\ CBAK\end{tabular}} & \textbf{\begin{tabular}[c]{@{}l@{}}pred.\\ COVL\end{tabular}} \\ \hline
		Base                  &  3.07                                                            &  95.2                                                            &  4.31                                                             & 3.43                                                              & 3.64  
		\\
		+ Multi-VQ
		& 3.12 &96.2 & 4.42 & 3.53 & 3.77                                                            
		\\ \hline
		+Wav2vec2.0           & 3.14                                                             & 96.3                                                             &  4.48                                                             &   3.58                                                            &   3.83                                                            \\
		+HuBERT               &  3.14                                                            & 96.4                                                             & 4.48                                                             & 3.59                                                              & 3.83                                                              \\
		+Data2vec             &  3.18                                                            &  96.7                                                            &   4.53                                                            &   3.64                                                            &   3.85                                                            \\ \hline
	\end{tabular}}
\end{table}	

\begin{table}[!t]
	\caption{The impact of VQ of different layers on the performance.}
	\label{tab:table3}
	\centering
	\resizebox{.98\columnwidth}{!}{
	\begin{tabular}{cccccc|cc}
		\hline
		\textbf{VQ0} & \textbf{VQ1} & \textbf{VQ2} & \textbf{VQ3} & \textbf{VQ4} & \textbf{VQ5} & \textbf{\begin{tabular}[c]{@{}c@{}}PESQ\\ (WB)\end{tabular}} & \textbf{\begin{tabular}[c]{@{}c@{}}STOI\\ (\%)\end{tabular}} \\ \hline
		\cmark        & \cmark        & \cmark        & \cmark        & \cmark        & \cmark        &   3.12                                                           & 96.2                                                             \\
		\xmark        & \cmark        & \cmark        & \cmark        & \cmark        & \cmark        &  3.09                                                            & 95.6                                                             \\
		\cmark        & \xmark        & \cmark        & \cmark        & \cmark        & \cmark        & 3.12                                                             &  96.0                                                            \\
		\cmark        & \cmark        & \xmark        & \cmark        & \cmark        & \cmark        & 3.11                                                             & 96.0                                                             \\
		\cmark        & \cmark        & \cmark        & \xmark        & \cmark        & \cmark        &  3.10                                                            & 95.9                                                             \\
		\cmark        & \cmark        & \cmark        & \cmark        & \xmark        & \cmark        & 3.10                                                            &  95.8                                                            \\
		\cmark        & \cmark        & \cmark        & \cmark        & \cmark        & \xmark        &  3.10                                                            &  95.8                                                            \\ \hline
	\end{tabular}}
\end{table}
\begin{figure}[!t]
	\centering
	\includegraphics[width=0.46\textwidth]{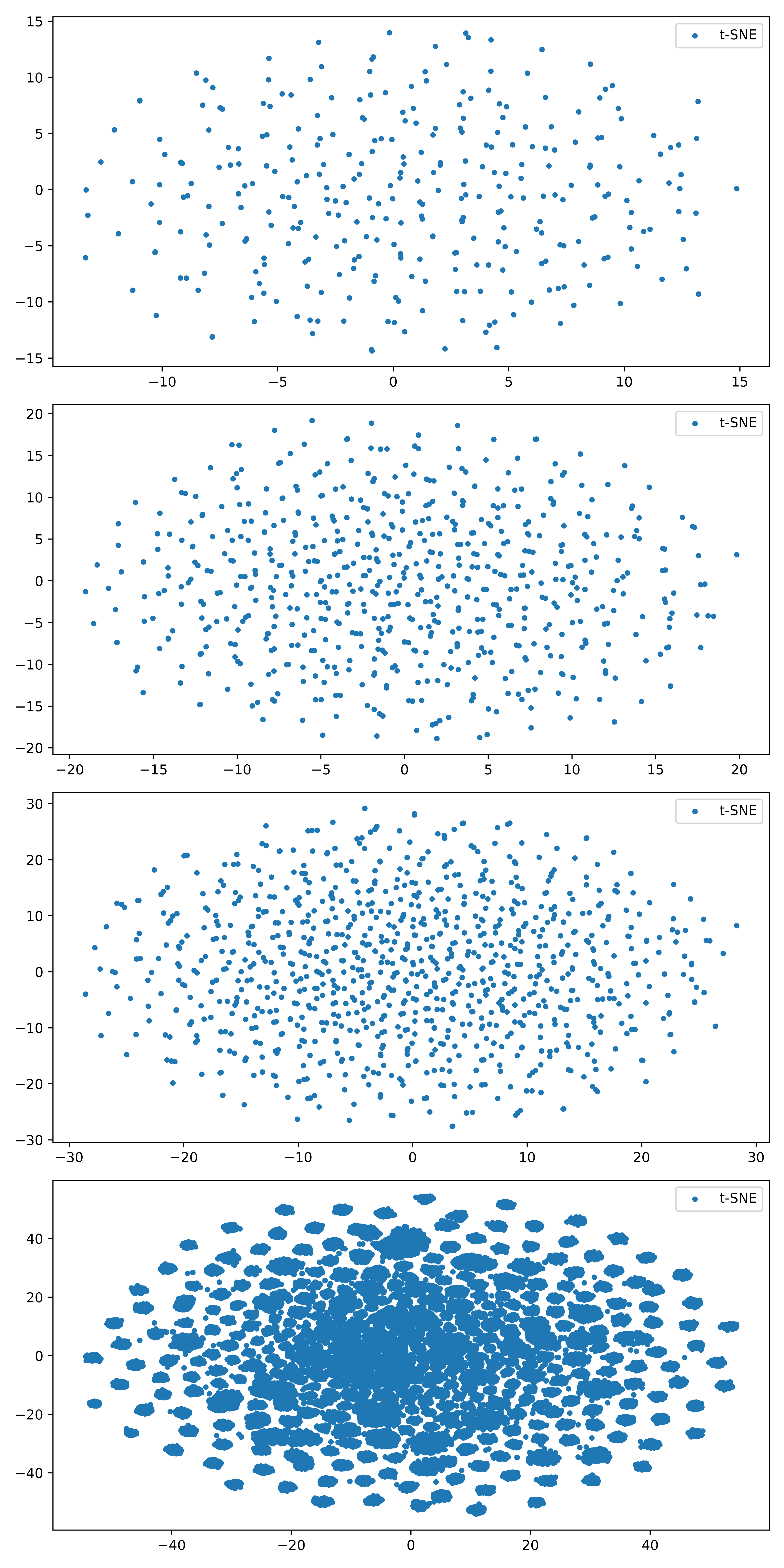}
	\caption{Codebook visualization with different number of learnable vectors, where a single codebook with 320, 640 and 960 learnable vectors and two codebooks with 320 learnable vectors are included from top to bottom.}
	\vspace{-0.2cm}
	\label{fig:fig3}
\end{figure}

Further, we experimentally analyze the impact of the VQ of different layers  in Table~\ref{tab:table3}.
It is clear that in case the VQ module is included  for each layer, the best SE performance is achieved.
The removal of the VQ at any layer leads to a performance decrease, meaning the necessity of the proposed multi-granularity VQ operations.
It seems that the removal of $VQ_0$ has a more significant impact on the performance, indicating that the discretized contextual features are more beneficial for model denoising.
$VQ_1$ has the smallest impact among all VQ modules, and a higher-layer VQ roughly affects the performance more. 
Due to the limited size of this dataset, these results might not reflect the analysis comprehensively, so in the future we will evaluate the proposed method in a larger-sized dataset.

Finally, we visualize the learned 128-dimensional codebooks using t-SNE, which are reduced to two dimensions and shown in Figure.~\ref{fig:fig3}, where a single codebook with 320 ($VQ_1$), 640 ($VQ_2$) and 960 ($VQ_3$) learnable vectors and two codebooks with 320 learnable vectors (320$\times$320 vectors in total, i.e., $VQ_0$) are included from top to bottom. It is clear that more learnable vectors used in the VQ codebook, more fine-grained speech embeddings can be extracted.  From the perspective of t-SNE distributions, the speech embedding extracted by $VQ_0$ shows a clearer clustering property, which is useful for classification. On the other hand, the multiple embeddings are not spatially overlapped, resulting in the so-called multi-granularity speech features, they are somehow complementary to represent the speech domain. That is why the inclusion of each VQ module positively contributes to the performance in Table~\ref{tab:table3}.

\vspace{-0.2cm}
\section{Conclusion}
\label{sec:conclusion}
In this paper, we investigated the application of multi-granularity VQ and pre-trained models to speech enhancement.
It was shown that using different VQ modules at different layers can improve the SE performance compared to using a single-fixed VQ.
The fusion of contextual features extracted by the pre-trained model and local features extracted by the encoder  allows the model to better model global and local information.
Compared with the self-supervised pre-trained wav2vec2.0 and HuBERT models that are dedicated to speech representation learning, data2vec seems more effective for SE as it can learn representations from general data (e.g., speech, images, video).

\section{Acknowledgements}

This work was supported by the National Natural Science Foundation of
China (62101523), Hefei Municipal Natural Science Foundation (2022012)
and Fundamental Research Funds for the Central Universities.

\bibliographystyle{IEEEtran}
\bibliography{mybib}

\end{document}